\documentclass[prb,floatfix,twocolumn,showpacs,amsmath,amssymb,
superscriptaddress]{revtex4}
\usepackage{graphicx}
\usepackage{dcolumn}
\usepackage{bm}
\begin{document}

\title{From Luttinger liquid to Mott insulator: the correct low-energy 
description of the one-dimensional Hubbard model by an unbiased variational 
approach}
\author{Manuela Capello}
\affiliation{INFM-Democritos, National Simulation Center and International
School for Advanced Studies (SISSA), I-34014 Trieste, Italy}
\author{Federico Becca}
\affiliation{INFM-Democritos, National Simulation Center and International
School for Advanced Studies (SISSA), I-34014 Trieste, Italy}
\author{Seiji Yunoki}
\affiliation{INFM-Democritos, National Simulation Center and International
School for Advanced Studies (SISSA), I-34014 Trieste, Italy}
\author{Michele Fabrizio}
\affiliation{INFM-Democritos, National Simulation Center and International
School for Advanced Studies (SISSA), I-34014 Trieste, Italy}
\affiliation{The Abdus Salam International Center for Theoretical Physics 
(ICTP), P.O. Box 586, I-34014 Trieste, Italy}
\author{Sandro Sorella}
\affiliation{INFM-Democritos, National Simulation Center and International
School for Advanced Studies (SISSA), I-34014 Trieste, Italy}
\date{\today}
\begin{abstract}
We show that a particular class of variational wave functions
reproduces the low-energy properties of the Hubbard model in one dimension. 
Our approach generalizes to finite on-site Coulomb 
repulsion the fully-projected wave function proposed by Hellberg and Mele
[Phys. Rev. Lett. {\bf 67}, 2080 (1991)]
for describing the Luttinger-liquid behavior of the doped $t{-}J$ model.
Within our approach, the long-range Jastrow factor emerges from a careful
minimization of the energy, without assuming any parametric form for the
long-distance tail. Specifically, in the conducting phase of the Hubbard model
at finite hole doping, we obtain the correct power-law behavior of the
correlation functions, with the exponents predicted by the Tomonaga-Luttinger 
theory. By decreasing the doping, the insulating phase is reached with a  
continuous change of the small-$q$ part of the Jastrow factor. 
\end{abstract}

\pacs{74.20.Mn, 71.10.Fd, 71.10.Pm, 71.27.+a}
\maketitle

\section{Introduction}

Electrons confined in one-dimensional (1D) systems exhibit peculiar  
non-Fermi liquid properties, that are by now rather well understood.~\cite{emery,soliom,alexei}
Let us just briefly mention some of them which will be useful for what follows.  
Due to phase-space limitations, particle-hole 
excitations in single-band 1D models are exhausted by collective 
charge and spin modes, which are dynamically independent, 
realizing what is commonly referred to as {\sl spin-charge separation}. 
When these modes are gapless, they propagate as acoustic waves  
(zero-sounds), hence can be identified by two parameters, the sound velocity 
$u_i$ and a dimensionless stiffness $K_i$, $i=\rho$ and $i=\sigma$ for 
charge and spin modes, respectively. 
Sometimes however the interaction opens a gap either in one of these sectors, 
for instance the charge sector in the Hubbard model at half-filling, or in both of them, 
like in a spontaneously dimerized chain.

Besides spin-charge separation, another manifestation of the 1D 
non-Fermi liquid behavior is the power-law decay with anomalous exponents of 
all correlation functions, when both spin and charge modes are gapless, 
or just some of them, when one of the two modes is gapped. 
Although these exponents are generically non universal, 
they all can be expressed in terms of the above mentioned $K_\rho$ and $K_\sigma$.
Therefore the finite set of parameters $u_\rho$, $u_\sigma$, $K_\rho$ and 
$K_\sigma$ is sufficient to characterize completely the asymptotic low-energy 
behavior of 1D interacting electron models, similarly to the finite number of 
parameters which identify the low-energy behavior of Landau-Fermi liquids in 
higher dimensions.~\cite{nozieres} Indeed, just in analogy with Fermi liquids, this kind of 
1D universal behavior was named ``Luttinger liquid'' by 
Haldane.~\cite{haldane1,haldane2,haldane3} 

In the case of non interacting electrons  $K_\rho=K_\sigma=1$. 
If spin SU(2) symmetry is unbroken and the spin modes are gapless, $K_\sigma$ 
remains one as for free fermions 
even in the presence of interaction and only $K_\rho$ parametrizes the anomalous exponents. 
In particular $K_\rho$ is smaller than one for repulsive interaction, and greater than one 
otherwise.  When both charge and spin sectors are gapless, the asymptotic expressions 
of the charge and spin equal-time 
correlation functions are, apart from possible logarithmic corrections, 
\begin{eqnarray}
&& \langle n(x) n(0) \rangle \sim \frac{K_\rho}{(\pi x)^2} + A_1
\frac{\cos(2k_F x)}{x^{K_\rho+1}} \nonumber \\
&&~~~~~~~~~~~~~~~+ A_2
\frac{\cos(4k_F x)}{x^{4K_\rho}}, \label{nn} \\
&& \langle {\bf S}(x) \cdot {\bf S}(0) \rangle \sim \frac{1}{(\pi x)^2} + B
\frac{\cos(2k_F x)}{x^{K_\rho+1}}, \label{SS}
\end{eqnarray}
where $n(x)$ and ${\bf S}(x)$ are the charge and spin density operators at  
position $x$, $k_F$ is the Fermi momentum, $A_1$, $A_2$, and $B$ are model-dependent constants. 
Analogously the singlet (and triplet) pairing correlations behave as 
\begin{equation}
\langle \Delta^\dag(x) \Delta(0) \rangle \sim \frac{1}{x^{K_\rho^{-1}+1}},
\end{equation}
where $\Delta^\dag(x)$ creates a singlet (or triplet) pair at position $x$.
Finally, the non-Fermi liquid character of 1D interacting models 
shows up transparently in the momentum 
distribution function near the Fermi momentum:
\begin{equation}
n_k - n_{k_F} \sim -  {\rm sign}(k-k_F) |k-k_F|^\theta,
\end{equation}
where $\theta$ is again expressed in terms of $K_\rho$ through the relation 
$\theta = (K_\rho-K_\rho^{-1}-2)/4$. 
For any finite interaction $K_\rho \ne 1$, hence the momentum distribution 
function has a power-law singularity at the Fermi level, in contrast to
the finite jump characteristic of Fermi liquids.
It is worth mentioning that, for non-frustrated models with repulsive interaction, 
the insulating phase at half-filling is characterized by all correlation functions 
decaying exponentially to zero apart from the spin-spin ones, which still decay as 
a power-law, formally like in~(\ref{SS}) with $K_\rho=0$.  
The actual value of the parameter $K_\rho$ depends upon the particular microscopic model, 
through the form and the strength of the interaction as well as the electron doping. 
For simple Hamiltonians $K_\rho$ can be explicitly calculated by 
the Bethe ansatz~\cite{schulz} or by exact diagonalization on small systems.~\cite{ogata1}
 
Although analytical techniques, especially bosonization~\cite{alexei}, give
important insights into the low-energy properties of 1D systems, 
both in gapless and in gapped phases, yet they do not provide a simple representation 
of the ground-state wave function (WF). Even in those simple models which are solvable 
by Bethe ansatz, the actual ground state WF turns out to be very difficult to interpret
and to deal with.
In the Hubbard model, for instance, the ground state WF is very involved within the
Bethe ansatz formalism and only in the strong-coupling limit it is possible 
to obtain significant simplifications because of the explicit factorization 
of the WF into a charge and a spin part.~\cite{ogata2}
For this reason there have been many attempts aimed to find out approximate
variational WFs, which, from one side, could correctly reproduce the peculiar 
properties of 1D systems but, from the other side, could also give a 
transparent interpretation of their physical properties.~\cite{shiba1,shiba2,mele,ogata3,gros}    
For instance it has been shown that the strong-coupling limit of the 
Hubbard model, where the on-site Coulomb repulsion $U$ prohibits doubly occupied sites,
is well described by a projected Slater determinant.~\cite{mele}
In particular the large-$U$ conducting phase at finite hole doping 
has been found to be properly represented through a 
fully-projected Gutzwiller WF supplemented by a long-range density-density
Jastrow factor. The latter is a crucial ingredient which allows to recover the 
anomalous power-law behavior of the correlation functions.
Indeed any short-range Jastrow factor cannot reproduce 
the correct Luttinger-liquid behavior since it is unable to affect the 
low-energy physics. For instance, it is well known that the simple Gutzwiller
WF has a finite jump in the momentum distribution function,~\cite{metzner} which is
certainly incorrect in 1D. 

In this paper we generalize this approach to 
a finite Coulomb repulsion $U$, namely to the case where charge fluctuations 
are still allowed.  We show that even at finite $U$ it is  
possible to design a consistent WF, which can faithfully describe the evolution from 
the Luttinger-liquid behavior at finite hole doping to the Mott insulating phase 
at half-filling. Again the crucial ingredient turns out to be a   
density-density Jastrow factor applied to a simple Slater determinant.
More specifically, we apply this variational WF to analyze the single-band Hubbard model. 
Thanks to the recent improvements in the energy minimization schemes 
for correlated WFs,~\cite{sr} the long-range tail of the Jastrow factor 
can be determined very accurately, without imposing any parametric form, 
even if the change of its tail may contribute to a very tiny energy gain. 
We will show that the low-energy properties of the Hubbard model 
are correctly reproduced within this unbiased variational approach.

The paper is organized as follow: in Sec.~\ref{model} we introduce the model
and the physical quantities we are interested in and in Sec.~\ref{result}
we present and discuss the results.

\section{The model}\label{model}

As we previously mentioned, in this work we aim to design a variational WF 
which is equally accurate for both weak and strong Coulomb repulsion.  
For sake of simplicity, we will test the quality of this WF in the well-known 
single-band Hubbard model on a finite chain with $L$ sites, $N$ particles and 
periodic boundary conditions. The Hamiltonian is 
\begin{equation}\label{hamiltonian}
H= -t \sum_{\langle i,j \rangle, \sigma} c^\dag_{i,\sigma} c_{j,\sigma} + h.c.
+U \sum_i  n_{i,\uparrow} n_{i,\downarrow},
\end{equation}
where we use the standard notations in which $c_{i,\sigma}$ 
($c^{\dag}_{i,\sigma}$) destroys (creates) an electron with spin $\sigma$ 
at the site $i$ and $n_{i,\sigma}= c^{\dag}_{i,\sigma} c_{i,\sigma}$. 

\begin{figure}
\includegraphics[width=\columnwidth]{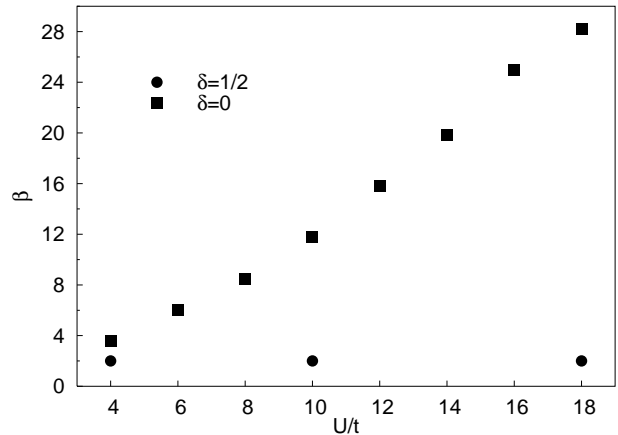}
\vspace{-0.5cm}
\caption{\label{fig1} 
The value of $\beta$ from Eq.~(\ref{RPA2}) for half-filling and 
quarter-filling for different ratios $U/t$.}
\end{figure}

\begin{figure}
\includegraphics[width=\columnwidth]{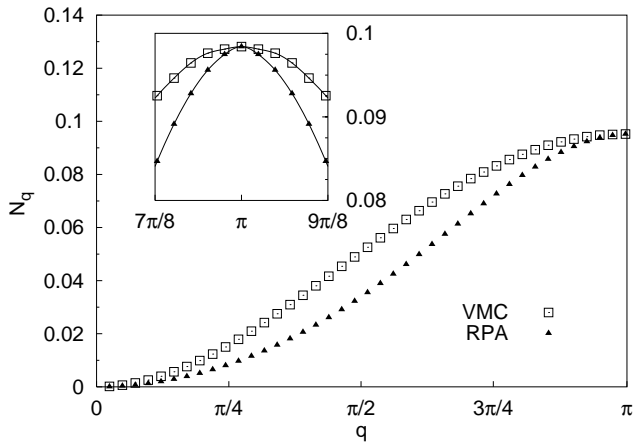}
\vspace{-0.5cm}
\caption{\label{fig2} 
Charge structure factor for $L=82$ and $U/t=10$ at half-filling: comparison
between the variational result and the RPA expression given by 
Eq.~(\ref{RPA2}). Inset: zoom around $q=\pi$. }
\end{figure}

The variational ansatz for the ground state WF is
\begin{equation}
|\Psi_N \rangle = {\cal J} |{\cal D} \rangle,
\label{wf}
\end{equation}
where ${\cal J}$ is a Jastrow factor and $|{\cal D} \rangle$ an uncorrelated 
Slater determinant, that, for simplicity, we assume to be the $N$-electron Fermi 
sea of the tight-binding model with dispersion $\epsilon_k = -2t \cos k$. 
Since spin SU(2) is unbroken, there is no need to introduce a spin-spin 
Jastrow factor, because a free-electron determinant already provides the 
correct value $K_\sigma=1$. Therefore, in order to preserve all the symmetries 
of the Hamiltonian, $\cal{J}$ is a purely density-density correlator 
written in term of the density 
operator $n_i=\sum_{\sigma}c^\dagger_{i,\sigma}c_{i,\sigma}$ as 
\[
{\cal J}=\exp\left[ -\frac{1}{2} \sum_{ij} v_{ij} n_i n_j\right],
\]
where, due to translation 
and inversion symmetry, $v_{i,j}$ can be described by using  $L/2$ independent
variational parameters $v(|i-j|)=v_{i,j}$; in particular, $v(L/2)$ can be set 
to zero since, due to the conservation of the total number of particles, 
$v_{ij} \to v_{ij} + {\rm const}$ provides only an irrelevant normalization
factor in the WF. 
 
A more sophisticated possibility would have been to choose $|{\cal D} \rangle$ as the 
ground state of a mean-field BCS Hamiltonian projected onto a fixed number of particles. 
For the simple Hubbard model of Eq.~(\ref{hamiltonian}), the latter ansatz 
leads to a slightly lower energy, without modifying the low-energy properties,
and, therefore, it will not be considered in the following. 
It should be mentioned, however, that in presence of a large 
next-nearest-neighbor hopping term, a gap in the BCS mean field Hamiltonian 
may open, leading to translational symmetry breaking at half-filling
 or to a phase of singlet pairs with dominant superconducting 
correlations at finite doping.~\cite{capello,fabrizio}

Summarizing, all the variational parameters are contained in the Jastrow 
coefficients $v(|i-j|)$ for all the $L/2$ independent distances in real space
and are calculated by a full energy minimization, without assuming any particular 
parametric form.

\begin{figure}
\includegraphics[width=\columnwidth]{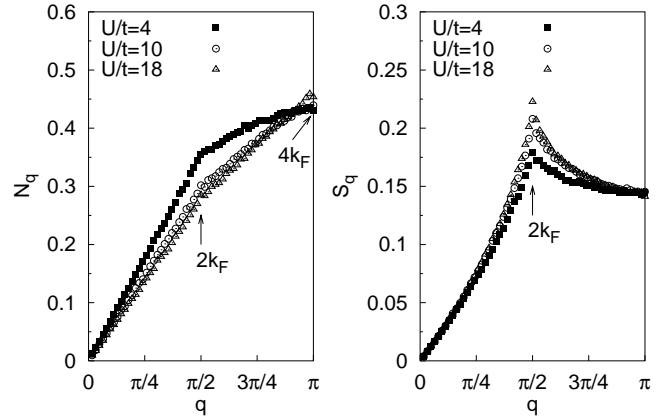}
\vspace{-0.5cm}
\caption{\label{fig3} 
Left panel: charge structure factor $N_q$ for the Hubbard model at
quarter-filling for $L=124$ and different $U/t$. 
Right panel: the same for the spin structure factor $S_q$.}
\end{figure}

Important quantities to assess the nature of the ground state are the 
static structure factor for the charge:
\begin{equation}
N_q = \frac{\langle \Psi_N| n_{-q}n_{q} |\Psi_N \rangle}
{\langle \Psi_N| \Psi_N \rangle},
\end{equation} 
and  similarly  for the spin:
\begin{equation}
S_q = \frac{\langle \Psi_N| S^z_{-q}S^z_{q} |\Psi_N \rangle}
{\langle \Psi_N| \Psi_N \rangle},
\end{equation} 
where $n_q$ and $S^z_q$ are the Fourier transform of the local density and
spin operators.

Another quantity that gives information on the WF is the quasiparticle weight:
\begin{equation}
Z_k = \frac{|\langle \Psi_{N-1}|  c_{k,\sigma}  |\Psi_N \rangle|^2}
{\langle \Psi_N| \Psi_N \rangle \langle \Psi_{N-1}| \Psi_{N-1} \rangle},
\label{zeta}
\end{equation}
where $|\Psi_N \rangle$ and $|\Psi_{N-1} \rangle$ are the WFs with $N$ and 
$(N-1)$ particles, $c_{k,\sigma}$ is the annihilation operator of a particle of 
momentum $k$ and spin $\sigma$. The $(N-1)$-particle state is obtained by  
the $N$-particle one by removing an electron from the Slater determinant, i.e.,
$|\Psi_{N-1} \rangle = {\cal J}\, c_{k,\sigma}\, |{\cal D} \rangle$.
In a Fermi liquid $Z_k$ is finite in the thermodynamic limit signaling 
the existence of coherent quasiparticles. On the contrary a non-Fermi liquid phase 
without quasiparticles is identified by a vanishing $Z_k$.

\begin{figure}
\includegraphics[width=\columnwidth]{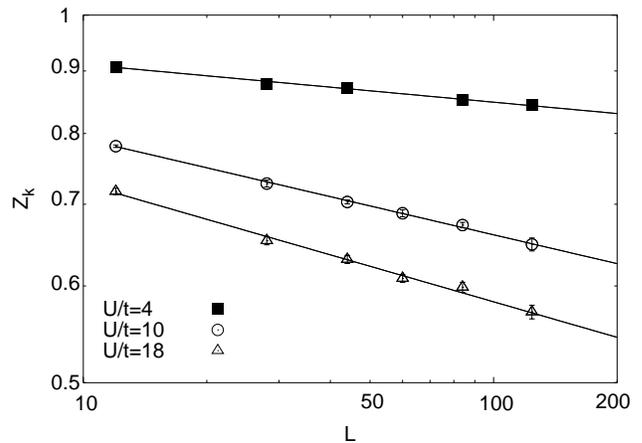}
\vspace{-0.5cm}
\caption{\label{fig4} 
Log-log plot of the quasiparticle weight $Z_k$ at $k=k_F=\pi/4$ as a function 
of $L$ in the quarter-filled Hubbard model for different values of $U/t$.
Lines are power-law fits.}
\end{figure}

\section{Results and discussion}\label{result}

The important role of the Jastrow factor has been already discussed by 
Hellberg and Mele in the context of the 1D $t{-}J$ model.~\cite{mele} 
In this case, it is possible to show analytically that the momentum 
distribution function of the variational WF has an algebraic singularity at 
$k_F$, with an exponent related to the strength of the Jastrow 
factor.~\cite{horsch}
Recently, we have shown~\cite{capello} that in the 1D Hubbard 
model a suitable long-range Jastrow can also drive a metallic Slater determinant 
into an insulating
phase and that, in the limit of small momenta, the Fourier transform of the 
Jastrow parameters $v_q$ are related to the charge-structure factor.
In particular, in the conducting regime and and for small momenta 
a random-phase-approximation (RPA) expression holds:~\cite{reatto}
\begin{equation}
N_q \sim \frac{N^0_q}{1+2 v_q N^0_q},
\label{RPA}
\end{equation}
where we indicate with $N^0_q$ the charge structure factor of the WF without 
${\cal J}$. For a conducting Slater determinant, like the free-electron state,
$N^0_q = |q|/\pi$ for small $q$. For a non-vanishing hole doping $\delta$ and
interaction $U/t$, by optimizing the variational WF, we obtain that the Jastrow factor is 
singular, i.e., $v_q \sim 1/|q|$. This fact is crucial to recover the 
correct low-$q$ behavior of $N_q$, whose linear slope is renormalized by the
interaction, leading to $N_q \sim K_\rho |q|/\pi$.
Therefore, the Jastrow factor has to be intrinsically long-range.
According to the RPA expression~(\ref{RPA}), a more singular Jastrow term 
$v_q \sim 1/q^2$ is needed to bring the system into the insulating 
phase at half-filling.~\cite{capello} 
In 1D, such a Jastrow factor leads to a confined phase, where perturbation 
theory does not apply, and, therefore, also the RPA is expected to fail.
Nonetheless, it turns out that, also within the insulating phase, the small-$q$ 
behavior of $N(q)$ is qualitatively reproduced by Eq.~(\ref{RPA}) 
and a more accurate empirical expression is given by:
\begin{equation}
N_q \sim \frac{N^0_q}{1+\beta v_q N^0_q},
\label{RPA2}
\end{equation}
where $\beta > 2$ is a constant that strongly
depends upon the electronic interaction. In Fig.~\ref{fig1}, we report
$\beta$ for different values of the ratio $U/t$ at half-filling, and, for
comparison, also at quarter-filling, where the value $\beta=2$ is recovered,
according to Eq.~(\ref{RPA}).
In the half-filled insulating phase, although for $q \to 0$ we have that 
$N_q \sim q^2$, the coefficient of the quadratic term is not simply related to the
Jastrow factor.
In this case, it is important to emphasize that the presence of the 
singular Jastrow factor $v_q \sim 1/q^2$ determines a qualitative change 
of the static structure factor $N(q)$ even at large momenta 
$q \sim 2 k_F = \pi$.
In fact, the charge structure factor $N^0(q)$ for the free Fermi gas is 
characterized by a cusp at $q \sim 2 k_F$, which is responsible of the 
well-known Friedel oscillations in a metal and the RPA expression cannot
remove this feature.
On the other hand, a true insulating phase does not possess this singularity
and, indeed, as shown in Fig.~\ref{fig2}, the charge structure factor 
$N(q)$ for the WF containing a singular Jastrow factor shows a smooth 
behavior around $q \sim \pi$.
This clearly indicates the non-perturbative and highly 
non-trivial effect implied by the formation of a confined state  
in this correlated WF between empty sites (holons) and doubly occupied 
ones (doublons). 

\begin{figure}
\includegraphics[width=\columnwidth]{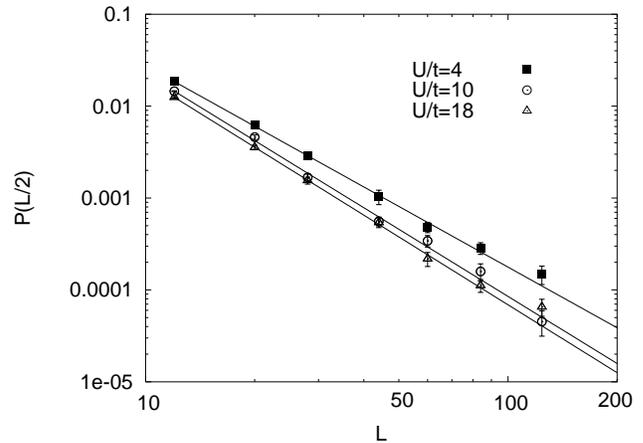}
\vspace{-0.5cm}
\caption{\label{fig5} 
Log-log plot of the pairing correlation function at the maximum distance
$P(L/2)$ as a function of $L$ in the quarter-filled Hubbard model for 
different values of $U/t$. Lines are power-law fits.}
\end{figure}

In order to demonstrate that the WF~(\ref{wf}) is able to capture the 
Luttinger-liquid metallic properties, 
we consider the quarter-filling case.
In Fig.~\ref{fig3}, we show the charge and spin structure factor for 
different values of $U/t$. For small momenta, the linear slope of $N_q$ is 
renormalized with respect to the non-interacting value, leading to 
$N_q \sim K_\rho |q|/\pi$. On the other hand, the small-$q$ behavior of $S_q$ 
is not affected by the interaction and we have that $S_q \sim |q|/4\pi$.
Notice that, in the presence of a strong interaction, the two singularities
at $2k_F$ and $4k_F$ are clearly visible in $N_q$, whereas in $S_q$,
only the singularity at $2k_F$ can be detected.
From the small-$q$ linear part of $N_q$, it is possible to extract the 
value of $K_\rho$ (see Table~\ref{tab1}), which is in very good agreement 
with the exact one.~\cite{schulz}
It is important to emphasize that the relationships among exponents of different 
correlation functions are correctly reproduced by our variational WF. 
Indeed, we can compare the value of the exponent $\theta$ found from a direct
evaluation of the quasiparticle weight~(\ref{zeta}) at $k=k_F$, i.e., 
$Z_k \sim 1/L^\theta$ (see Fig.~\ref{fig4}), with the one obtained with 
$\theta=(K_\rho+K_\rho^{-1}-2)/4$ by using the value of $K_\rho$ extracted 
from the linear slope of $N_q$. 
As reported in Table~\ref{tab1}, we obtain an excellent agreement for the 
values of the interaction $U/t$ considered.
Finally, we can also calculate the singlet pairing correlations
\begin{equation}
P(r) = \frac{1}{L} \sum_i \langle \Psi_N| \Delta_{i+r} \Delta^\dag_{i} 
|\Psi_N \rangle,
\end{equation}
where
\begin{equation}
\Delta^\dag_{i} = c^\dag_{i,\uparrow} c^\dag_{i+1,\downarrow} -
c^\dag_{i,\downarrow} c^\dag_{i+1,\uparrow}
\end{equation}
creates a singlet pair of electrons at nearest neighbors. In order to
calculate the exponent $\alpha$ related to the decay of 
$P(r) \sim 1/r^\alpha$, we consider the pairing correlation at the maximum
distance $P(L/2)$ for different sizes, see Fig.~\ref{fig5}.
In this case, the signal is very small and a precise determination of the 
critical exponent is quite difficult. Nonetheless, the results reported 
in Table~\ref{tab1} are rather satisfactory and not too far from the ones
obtained with the exact relation $\alpha=K_\rho^{-1}+1$.~\cite{schulz}

\begin{table}
\caption{\label{tab1} 
Critical exponents for the 1D Hubbard model at quarter-filling:
$K_\rho$ is found from the low-$q$ behavior of $N_q$, 
$\theta_c=(K_\rho+K_\rho^{-1}-2)/4$,
and $\theta$ is found by fitting $Z_k$ with $Z_k \sim 1/L^\theta$.
The last two columns refer to the critical exponent of the pairing 
correlations: $\alpha$ is found from the pairing correlation at the maximum
distance $P(L/2) \sim 1/L^\alpha$ and $\alpha_c=K_\rho^{-1}+1$. 
In the first column, we report the exact value of $K_\rho$.}
\begin{tabular}{ccccccc}
\hline \hline
$U/t$ & $K_\rho^{exact}$ & $K_\rho$ & $\theta$ & $\theta_c$ & $\alpha$ & $\alpha_c$ \\
\hline \hline
4     & 0.711            & 0.705(3) & 0.031(5) & 0.031(3)   & 2.1(1)   & 2.42(6) \\
10    & 0.594            & 0.595(3) & 0.078(5) & 0.072(3)   & 2.4(1)   & 2.68(9) \\
18    & 0.551            & 0.550(3) & 0.097(5) & 0.092(3)   & 2.5(2)   & 2.82(9) \\
\hline \hline
\end{tabular}
\end{table}

\begin{figure}
\includegraphics[width=\columnwidth]{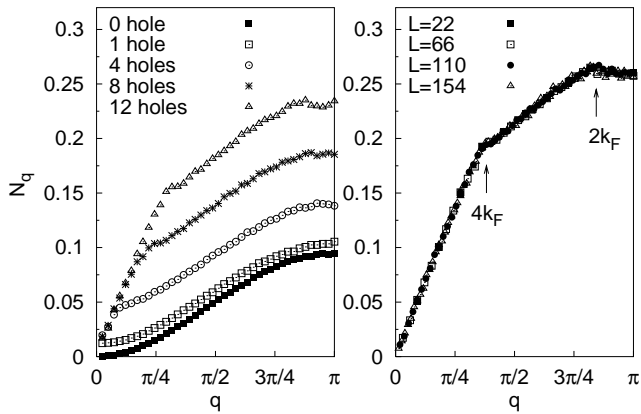}
\vspace{-0.5cm}
\caption{\label{fig6} 
Left panel: charge structure factor $N_q$ for $L=82$, $U/t=10$, and
different hole dopings. Right panel: $N_q$ for $\delta=2/11$ and different 
$L$.}
\end{figure}

Let us now consider how the insulating case is reached by decreasing the hole 
concentration. First of all, it should be mentioned that, not too close to the
insulating phase at half-filling, the charge and spin structure factor have small 
size effects and, therefore, reliable calculations are possible even without using too 
large $L$. As an example, we report in Fig.~\ref{fig6} (right panel), the case of doping  
$\delta=2/11$, where we can see that there are no appreciable differences 
in $N_q$ from $L=22$ to $L=154$.

In the doped region, the system is always conducting, 
$N_q$ having a linear behavior for small momenta, with a slope that depends 
upon $U/t$ and $\delta$. For sufficiently small hole doping, it turns out that 
the linear regime is limited to a small window around $q=0$, whereas for 
larger momenta, $N_q$ acquires a finite curvature, see Fig.~\ref{fig6}
(left panel). The two different regimes are separated by the singularity at 
$q=4k_F=2\pi \delta$, and, therefore, by decreasing $\delta$, the width of 
the linear regime shrinks, the slope being almost constant.
Therefore, we arrive at the empirical result:
\begin{equation}
N_q \sim \frac{K_\rho |q|}{\pi} \Theta(4k_F-q) + 
(c+q^2) \Theta(q-4k_F),
\label{peculiarnq}
\end{equation}
where $k_F=(1-\delta)\pi/2$, $\Theta(x)$ indicates the Heaviside step function 
and $c$ is a constant determined by imposing continuity of $N_q$ at $q=4k_F$.
This singular behavior, with the kink at $q=4k_F$, is entirely due to 
correlation and it is compatible with the exact result that $K_\rho$ remains 
finite, more precisely $K_\rho \to 1/2$, for $\delta \to 0$.~\cite{schulz}
It is important to stress that, by approaching half-filling, the width of
the linear regime shrinks with doping and reduces to zero for $\delta \to 0$.

\begin{figure}
\includegraphics[width=\columnwidth]{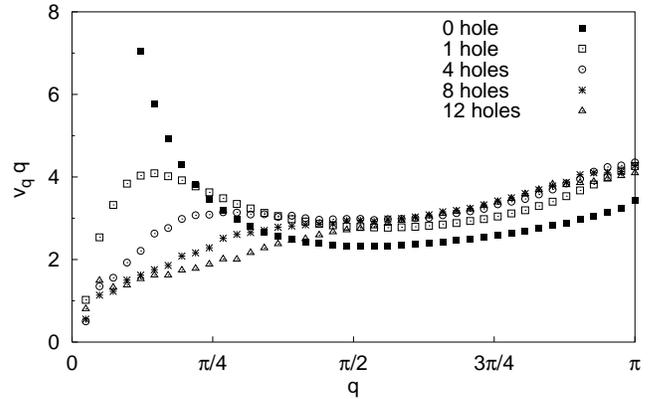}
\vspace{-0.5cm}
\caption{\label{fig7} 
Jastrow factor $v_q$ (multiplied by $q$) for different hole dopings, 
obtained by a careful minimization of the energy for $U/t=10$ and $L=82$.}
\end{figure}

\begin{figure}
\includegraphics[width=\columnwidth]{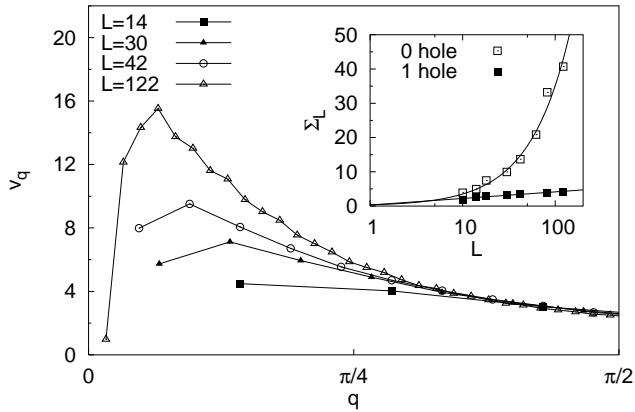}
\vspace{-0.5cm}
\caption{\label{fig8} 
Jastrow factor $v_q$ for one hole, $U/t=18$ and different $L$. Lines connect the points
Inset: the quantity $\Sigma_L = 1/L \sum_{q \ne 0} v_q$ (see text) as a function of $L$ 
for half-filling and one hole. Lines are fits of the data.}
\end{figure}

For completeness, we report in Fig.~\ref{fig7} the form of the Jastrow 
factor at half-filling and for small concentration of holes considered
in Fig.~\ref{fig6}.
Starting from the insulating phase, upon doping, $v_q$ moves away from 
$v_q \sim 1/q^2$, and becomes less singular, i.e., $v_q \sim 1/|q|$. 
Clearly, at very small doping, the size effects affect the small-$q$ part,
and, in particular, for the smallest momentum we can have some deviations
from the expected $v_q \sim 1/|q|$ behavior. 

\begin{figure}
\includegraphics[width=\columnwidth]{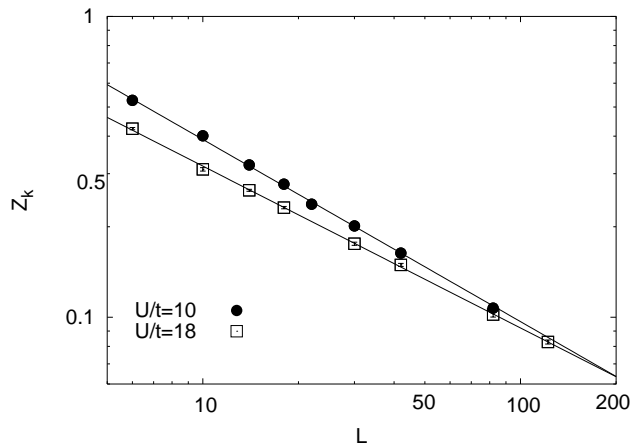}
\vspace{-0.5cm}
\caption{\label{fig9} 
Log-log plot of $Z_k$ as a function of $L$ for $U/t=10$ and $18$. 
Lines are power-law fits.}
\end{figure}

The correct minimization of the Jastrow factor is particularly important
for having an accurate description of $Z_k$, especially when approaching
half-filling. Indeed, in this case, the Jastrow factor for one hole is 
considerably different from the insulating one for small $q$'s 
(see Fig.~\ref{fig8}), and one has to optimize both WFs with 
$N$ and $(N-1)$ particles.~\cite{note}
This difference can be appreciated by considering 
$\Sigma_L = 1/L \sum_{q \ne 0} v_q$, which diverges linearly with
the system size $L$ if $v_q \sim 1/q^2$ and, instead, diverges only 
logarithmically if $v_q \sim 1/q$. In Fig.~\ref{fig8}, we report $\Sigma_L$
as a function of $L$ for the insulating state and for the one-hole case:
the difference between the two cases clearly demonstrates the different
behavior of $v_q$ for small momenta.
By a careful minimization of both the WFs, it is possible to recover
the result that $\theta=1/2$ independently of $U/t$.~\cite{parola}
Indeed, upon increasing $U/t$, our variational WF gives a rather 
accurate description of the insulating phase, the size effects being 
strongly reduced due to the small correlation length expected at large $U/t$.
In this limit, we obtain  a reasonable good agreement with the exact 
exponent for the quasiparticle weight (see Fig.~\ref{fig9}):
$\theta=0.60 \pm 0.05$ and $\theta=0.55 \pm 0.05$ 
for $U/t=10$ and $U/t=18$, respectively.
On the other hand, it should be mentioned that a naive calculation with a 
singular Jastrow $v_q \sim 1/q^2$ for both WFs would lead to a wrong 
exponential behavior of the quasiparticle weight.

In conclusion, we have shown that an accurate description of the charge
and spin properties of the 1D Hubbard model is possible within the simple 
variational ansatz given by Eq.~(\ref{wf}). The main ingredient of the 
variational WF is represented by a singular density-density Jastrow factor, 
whose long-range part determines the low-energy behavior of the correlation 
functions and can be found by optimization of the energy.
In the conducting regime, the small-$q$ Fourier transform of the 
density-density Jastrow term diverges like $v_q \sim 1/|q|$, implying the 
correct renormalization of the charge structure factor for small momenta, 
i.e., $N_q \sim K_\rho|q|/\pi$. By approaching the insulating phase and 
by decreasing the hole doping, the Jastrow factor modifies its small-$q$ 
part and, eventually, it acquires a more singular behavior like 
$v_q \sim 1/q^2$.
Within this context, it is possible to obtain a consistent scenario, where 
we recover the correct relationships among the exponents of different physical 
quantities. Obviously, our approach works due to the quality of the variational 
ansatz in 1D. Despite the difficulty to define very accurate WFs in higher 
spatial dimensions, the possibility to obtain the correct low-energy 
description of a Hamiltonian by using a simple optimization of the energy 
is very appealing and work is in progress for the generalization of this ansatz 
to two-dimensional systems, where no exact solutions are available. 

We acknowledge the constant and fruitful interaction with E. Tosatti, who
gave us an important contribution to our work. We thank A. Parola for 
stimulating discussions and for having provided us the exact values of 
$K_\rho$. This research has been supported by PRIN-COFIN 2004 and INFM.


\begin{thebibliography}{99}

\bibitem{emery} See for instance, V.J. Emery, in {\it Highly conducting
   one-dimensional solids}, Ed. by J.T. Devreese, R.P. Evrard, and 
   V.E. van Doren (Plenum, New York, 1979), p.247.
\bibitem{soliom} J. Solyom, Adv. Phys. {\bf 28}, 209 (1979).
\bibitem{alexei} See for instance  A.O. Gogolin, A.A. Nersesyan, and A.M. Tsvelik,
{\it Bosonization and Strongly Correlated Systems} (Cambridge Univ. Press,
Cambridge, 1999).
\bibitem{nozieres} See for instance, P. Nozieres and D. Pines, 
   {\it Theory of Quantum liquids} (Perseus, Cambridge 1999).
\bibitem{haldane1} F.D.M. Haldane, J. Phys. C {\bf 14}, 2585 (1981).
\bibitem{haldane2} F.D.M. Haldane, \prl {\bf 45}, 1358 (1980).
\bibitem{haldane3} F.D.M. Haldane, \prl {\bf 47}, 1840 (1981).
\bibitem{schulz} H.J. Schulz, Int. J. Mod. Phys. B {\bf 5}, 57 (1991).
\bibitem{ogata1} M. Ogata, M.U. Luchini, S. Sorella, and F.F. Assaad, 
   \prl {\bf 66}, 2388 (1991).
\bibitem{ogata2} M. Ogata and H. Shiba, \prb {\bf 41}, 2326 (1990).
\bibitem{shiba1} H. Yokoyama and H. Shiba, J. Phys. Soc. Jpn. {\bf 56}, 
   1490 (1987)
\bibitem{shiba2} H. Yokoyama and H. Shiba, J. Phys. Soc. Jpn. {\bf 59}, 
   3669 (1990)
\bibitem{mele} C.S. Hellberg and E.J. Mele, \prl {\bf 67}, 2080 (1991). 
\bibitem{ogata3} H. Yokoyama and M. Ogata, \prl {\bf 67}, 3610 (1991)
\bibitem{gros} C. Gros and R. Valenti, Mod. Phys. Lett. B {\bf 7}, 119 (1993).
\bibitem{metzner} W. Metzner and D. Vollhardt, \prl {\bf 59}, 121 (1987); 
   \prb {\bf 37}, 7382 (1988).
\bibitem{sr} S. Sorella, \prl {\bf 80}, 4558 (1998); \prb {\bf 64}, 
   024512 (2001).
\bibitem{capello} M. Capello, F. Becca, M. Fabrizio, S. Sorella, 
   and E. Tosatti, \prl {\bf 94}, 026406 (2005).
\bibitem{fabrizio} M. Fabrizio, \prb {\bf 54}, 10054 (1996).
\bibitem{horsch} N. Kawakami and P. Horsch, \prl {\bf 68}, 3110 (1992);
   C.S. Hellberg and E.J. Mele, \prl {\bf 68}, 3111 (1992).
\bibitem{reatto} L. Reatto and G.V. Chester, Phys. Rev. {\bf 155}, 88 (1967).
\bibitem{note} Deep in the conducting phase, the Jastrow factor does not 
   change much from $N$ and $(N-1)$ particles, and the optimization of the 
   two WFs is not mandatory.
\bibitem{parola} S. Sorella and A. Parola, \prb {\bf 57}, 6444 (1998).
\end{thebibliography}
\end{document}